\documentclass[10pt,twocolumn,letterpaper]{article}

\usepackage[pagenumbers]{cvpr} 

\usepackage[dvipsnames]{xcolor}

\usepackage{multirow}
\usepackage{makecell}

\newlength\savedwidth

\definecolor{cvprblue}{rgb}{0.21,0.49,0.74}
\usepackage[pagebackref,breaklinks,colorlinks,citecolor=cvprblue]{hyperref}
\usepackage[accsupp]{axessibility} 

\usepackage{tikz}
\usepackage{gensymb}

\DeclareMathOperator*{\argmax}{arg\,max}

\def\eg{\emph{e.g}\onedot} 
\def\ie{\emph{i.e}\onedot}

\def\etal{\emph{et al}\onedot}

\def\paperID{} 
\def\confName{CVPR}
\def\confYear{2024}

\title{Learned Scanpaths Aid Blind Panoramic Video Quality Assessment}

\author{Kanglong Fan$^{1}$, Wen Wen$^{1}$, Mu Li$^{2}\thanks{Corresponding author.}$, Yifan Peng$^{3}$, and Kede Ma$^{1}$\\
$^1$ City University of Hong Kong, 
$^2$ Harbin Institute of Technology, Shenzhen\\
$^3$ The University of Hong Kong\\
{\tt  \{kanglofan2-c,wwen29-c\}@my.cityu.edu.hk, limuhit@gmail.com}\\
{\tt evanpeng@hku.hk, kede.ma@cityu.edu.hk}\\
\url{https://github.com/kalofan/AutoScanpathQA}
}

\begin{document}
\maketitle
\begin{abstract}

Panoramic videos have the advantage of providing an immersive and interactive viewing experience. Nevertheless, their spherical nature gives rise to various and uncertain user viewing behaviors, which poses significant challenges for panoramic video quality assessment (PVQA). 
In this work, we propose an end-to-end optimized, blind PVQA method with explicit modeling of user viewing patterns through visual scanpaths. Our method consists of two modules: a scanpath generator and a quality assessor. The scanpath generator is initially trained to predict future scanpaths by minimizing their expected code length and then jointly optimized with the quality assessor for quality prediction. Our blind PVQA method enables direct quality assessment of panoramic images by treating them as videos composed of identical frames. Experiments on three public panoramic image and video quality datasets, encompassing both synthetic and authentic distortions, validate the superiority of our blind PVQA model over existing methods.

\end{abstract}

\section{Introduction}
\label{sec:intro}
The rapid advancement of multimedia technologies has marked the beginning of a new era characterized by the proliferation of panoramic videos~\cite{ng2005data}. Such type of digital data offers an immersive and interactive viewing experience that is transforming the way we consume multimedia. Therefore, assessing and ensuring the visual quality of panoramic videos is increasingly important, as it shapes the users' viewing experience and the triumph of any product or service based on panoramic videos~\cite{yu2015framework}. Unlike their planar counterparts, panoramic videos provide a 360$^{\circ}$ broad view with a spherical data structure, which poses significant computational challenges for panoramic video quality assessment (PVQA). Moreover, the diverse and uncertain user viewing behaviors (in the form of visual scanpaths) induced by the spherical structure further complicate the quality prediction process. Addressing these challenges requires novel PVQA models that take into account both the spherical data structure as well as user viewing patterns.  

\begin{figure}[!tbp]
\scriptsize
\centering
\includegraphics[width=0.48\textwidth]{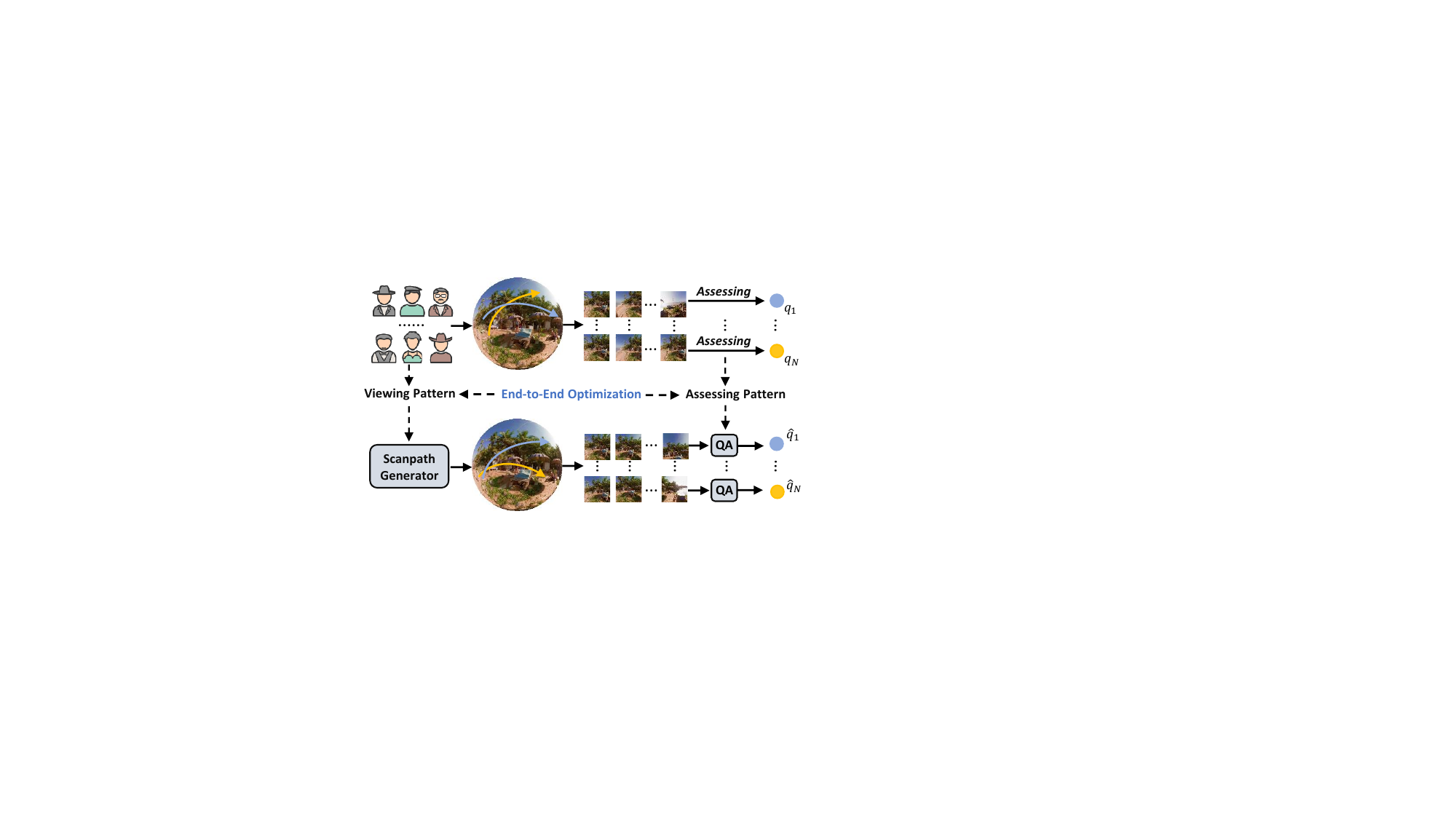}
\vspace{-6pt}
\caption{Analogy between human subjects and our end-to-end optimized method for panoramic video quality assessment.
}  
\vspace{-12pt}
\label{fig:1}
\end{figure}

In the quality assessment of panoramic images and videos, three approaches are commonly employed: sphere-to-plane projection onto a 2D plane, rectilinear projection onto multiple viewports, and
direct processing using spherical operations. According to the Theorema Egregium by Gauss, all sphere-to-plane map projections~\cite{zakharchenko2016quality,sun2017weighted,wen2024perceptual}
are impeded by non-uniform sampling and geometric distortions, which may bias subsequent \textit{planar} quality prediction. While spherical operators give a better account for the panoramic data structure, they are generally computationally prohibitive and, more importantly, may not faithfully reflect user viewing patterns~\cite{yu2015framework,chen2018spherical, yang2021panoramic}. To overcome these computational difficulties, several methods seek to sample and process rectilinear viewports~\cite{li2019viewport,xu2020blind,sui2021perceptual,fu2022adaptive, sui2023perceptual,wu2023assessor360}. Of particular interest are 
scanpath-based methods, which sample, along visual scanpaths~\cite{noton1971scanpaths, noton1971scanpathsineye}, sequences of rectilinear viewports at discrete time instances. This sampling process turns panoramic images and videos into moving-camera videos, amenable to \textit{planar} VQA.

 By closely imitating how humans perceive visual distortions in virtual environments (see Figure~\ref{fig:1}), scanpath-based methods~\cite{sui2021perceptual,sui2023perceptual,wu2023assessor360} have demonstrated remarkable efficacy in the quality of panoramic images. Nonetheless, some methods~\cite{sui2021perceptual} rely on human visual scanpaths for assessment, which are cumbersome and time-consuming to obtain, thus limiting their applications in fully-automated situations. Some other methods~\cite{sui2023perceptual, wu2023assessor360} design and refine the scanpath generator separately from the quality predictor, which is bound to be suboptimal. Moreover, while all methods prove effective with panoramic images, their adaptability for use with panoramic videos remains unclear.

In this work, we further pursue the scanpath-based methods for end-to-end optimized blind PVQA. Our method consists of two modules: a scanpath generator and a quality assessor. Our scanpath generator is probabilistic, which takes historical scanpaths as input
and is pre-trained to predict future scanpaths by minimizing their expected code length~\cite{li2023scanpath}. The scanpath generator and the quality assessor are then jointly optimized to explain human perceptual scores of panoramic videos. 
To enable end-to-end optimization, we employ the reparameterization trick~\cite{kingma2013auto,jang2017categorical} to allow differentiable scanpath sampling and adopt sub-gradients to handle discontinuities of the interpolation kernel~\cite{Jaderberg2015Spatial} for viewport sequence generation. 
Our blind PVQA method not only eliminates the need for human scanpaths, but also supplies a lightweight and differentiable scanpath generator that can work with any planar VQA model. Furthermore, our method is ``backward compatible,'' in the sense that it handles panoramic images with no modification. We test the proposed blind PVQA models on three public panoramic image and video quality datasets~\cite{wen2024perceptual,sun2018cviqd,duan2018perceptual}, covering both synthetic and authentic distortions. Under both in-dataset and cross-dataset settings, our models consistently exhibit better quality prediction performance. 

\begin{figure*}[!tbp]
\scriptsize
\centering
\includegraphics[width=1\textwidth]{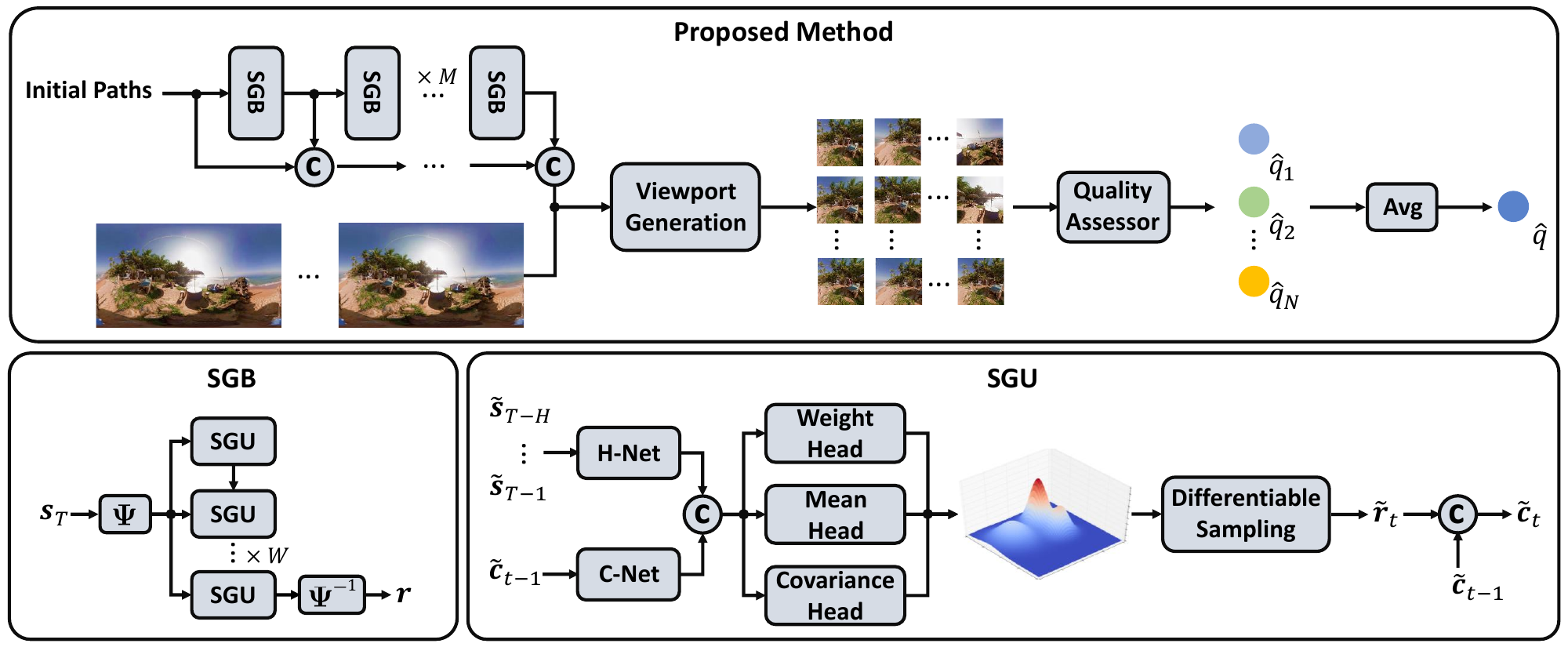}
\vspace{-6pt}
\caption{Overview of the proposed blind PVQA method, consisting of a scanpath generator and a quality assessor. The basic component of the scanpath generator is the scanpath generation unit (SGU), which utilizes the historical and causal relative scanpaths to produce the GMM parameters for differentiable sampling of the current viewpoint. By assembling $W$ SGUs, we create a scanpath generation block (SGB), which autoregressively predicts a future scanpath of $W$ viewpoints. We further stack $M$ SGBs to generate a long-term scanpath of $M \times W +H$ viewpoints, where $H$ is the length of the initial path. By adjusting the number of initial paths (denoted by $N$), we can sample
 $N$ scanpaths, along which we produce $N$ viewport sequences as  input to the quality assessor.  }
\vspace{-11pt}
\label{fig:2}
\end{figure*}

\section{Related Work}
\label{sec:related}
We review two highly relevant topics, scanpath generation and quality assessment of panoramic images and videos.

\subsection{Scanpath Generation}
Typical inputs to a panoramic scanpath generator include the saliency map, optical flow map, and historical scanpath. To improve saliency detection, Nguyen~\etal~\cite{nguyen2018your} compiled a panoramic video saliency dataset, while Xu~\etal~\cite{xu2018gaze} focused on relative viewpoint displacement prediction. Apart from the historical scanpath, Li~\etal~\cite{li2019very} incorporated ``future'' scanpaths from other users to facilitate cross-user transfer learning. Through an in-depth root-cause analysis, Rond\'{o}n~\etal~\cite{rondon2022track} discovered that visual features have a minimal impact on the prediction of short-term scanpaths (\eg, $\le 2$ seconds). Motivated by their findings, Chao~\etal~\cite{chao2021transformer} trained a Transformer~\cite{vaswani2017attention} to predict future scanpaths based solely on historical scanpaths. 

The above-mentioned methods~\cite{nguyen2018your, xu2018gaze, li2019very, rondon2022track, chao2021transformer} treat scanpath generation as a deterministic prediction task, neglecting the inherent scanpath diversity and uncertainty. 
As a departure, Li~\etal~\cite{li2023scanpath} formulated scanpath generation as a density estimation problem, which can be implemented by expected code length minimization. In our work, we adopt Li's approach~\cite{li2023scanpath} to learn multi-user viewing patterns and generate human-like scanpaths.

\subsection{Quality Assessment}
Current PVQA models are primarily derived from planar image and video quality methods, which are applied to three types of data representations: the projected 2D plane, spherical surface, and projected rectilinear viewport.

Planar domain methods~\cite{sun2017weighted,zakharchenko2016quality,wen2024perceptual} aim to rectify geometric distortions and mitigate uneven sampling that results from the sphere-to-plane projection. These include the latitude-adaptive weighting~\cite{sun2017weighted}, Craster parabolic projection~\cite{zakharchenko2016quality}, and pseudocylindrical representation~\cite{wen2024perceptual}. Spherical domain methods, such as S-PSNR~\cite{yu2015framework} and S-SSIM~\cite{chen2018spherical}, compute and pool local quality measurements over the sphere. Yang~\etal~\cite{yang2021panoramic} trained a non-local spherical neural network~\cite{cohen2018spherical,wang2018non} to extract spatiotemporal information from panoramic videos. Viewport domain methods prioritize the extraction of visually informative viewports for quality analysis. Li~\etal~\cite{li2019viewport} introduced a two-step approach that involves viewport proposal and quality assessment. Xu~\etal~\cite{xu2020blind} built a graph over the extracted viewports, and Fu~\etal~\cite{fu2022adaptive} constructed hypergraphs to represent the semantic interactions between viewports. One limitation of current viewport proposal methods is that they do not accurately reflect the human viewing experience.

Sui~\etal~\cite{sui2021perceptual} pioneered scanpath-based methods for PVQA, under the category of viewport domain methods. To eliminate the dependency on human scanpaths, Sui~\etal~\cite{sui2023perceptual} adopted a deep Markov model~\cite{sui2023scandmm} to generate scanpaths. Meanwhile, Wu~\etal~\cite{wu2023assessor360} handcrafted a simple scanpath generator based on the entropy feature and equator bias. These methods are tailored for panoramic images and are not end-to-end optimized. In contrast, we aim ambitiously for an end-to-end optimized quality assessment method for panoramic videos, with the added benefit of being backward compatible with panoramic images.

\section{Proposed Method}
As illustrated in Figure~\ref{fig:2}, our method consists of two modules: a scanpath generator and a quality assessor. Given a panoramic video, we first specify a starting point $(\phi_0,\theta_0)$, a viewing duration $S$, and $N$ initial paths. The scanpath generator autoregressively samples $N$ scanpaths based on the initial and already generated path segments. Along these scanpaths, we apply a differentiable viewport generation technique to extract $N$ viewport sequences from the input panoramic video. Each viewport sequence (as a planar video) is fed to the quality assessor, whose predicted score is subsequently aggregated into an overall quality estimate of the panoramic video. 

\subsection{Scanpath Generator}
\noindent\textbf{Probabilistic Scanpath Modeling}.
To capture the uncertainty and diversity of human scanpaths, we formulate panoramic scanpath generation as a density estimation problem:
\begin{align}
    \max p(\boldsymbol{r}\vert\boldsymbol{s}),
\end{align} 
\noindent
where $\boldsymbol{s} = \{(\phi_0,\theta_0), \ldots, (\phi_t,\theta_t),\ldots,(\phi_{T-1},\theta_{T-1})\}$ is the historical scanpath as the condition and $\boldsymbol{r}= \{ (\phi_T,\theta_T), \ldots, (\phi_{T+W-1},\theta_{T+W-1})\}$ is the future scanpath to be predicted. Herein, $W$ is the prediction horizon, and $(\phi_t,\theta_t)$ is the $t$-th viewpoint in the Euler coordinate system. Mathematically, $p(\boldsymbol{r}\vert\boldsymbol{s})$ can be decomposed as
\begin{align}\label{eq:chain}
p(\boldsymbol{r}\vert \boldsymbol{s}) =\prod_{t=0}^{W-1} p\left({\phi}_{T+t},{\theta}_{T+t} \Big\vert \boldsymbol{s},\boldsymbol c_{t} \right) ,
\end{align}
where $\boldsymbol c_{t} = \{({\phi}_{T},{\theta}_{T}), \ldots, ({\phi}_{T+t-1},{\theta}_{T+t-1})\}$ is referred to as the causal path that includes all estimated viewpoints before $({\phi}_{T+t},{\theta}_{T+t})$, and $\boldsymbol c_{0} = \emptyset$. The chain rule  suggests estimating the conditional probability $p\left({\phi}_{T+t},{\theta}_{T+t} \Big\vert \boldsymbol{s},\boldsymbol c_{t} \right)$ autoregressively. We further make the Markovian assumption: prediction of the current viewpoint is conditionally independent of viewpoints that are temporally further distant, given the most recent $H$ viewpoints.
This leads to a truncated historical path context $\boldsymbol{s}_{T}=\{ (\phi_{T-H},\theta_{T-H}), \ldots, (\phi_{T-1},\theta_{T-1})\}$. 

We parameterize  the  probability $p\left(\boldsymbol{r}_t \Big\vert \boldsymbol{s}_T,\boldsymbol c_{t} \right)$, where  $\boldsymbol{r}_t = (\phi_{T+t},\theta_{T+t})$, by a Gaussian mixture model (GMM) with $K$ components:
\begin{align}\label{eq:GMMcp}
     p\left(\boldsymbol{r}_{t} \Big\vert \boldsymbol{s}_T,\boldsymbol c_{t} \right) = 
     \sum_{i=1}^K\alpha_i \mathcal{N}_i(\boldsymbol{r}_t;\boldsymbol \mu_i, \boldsymbol \Sigma_i) ,
\end{align}
where  $\alpha_i$ is the $i$-th mixture weight, $\boldsymbol \mu_i$ and $\boldsymbol \Sigma_i$ represent the mean vector and the covariance matrix of the $i$-th Gaussian component, respectively. This parametrization can be straightforwardly done by training a density estimation network for parameter estimation. As illustrated in Figure~\ref{fig:2}, this network is inside the scanpath generation unit (SGU) and is composed of two subnetworks to process the historical path context  $\boldsymbol{s}_T$ and the causal path context $\boldsymbol{c}_t$, which we denote by H-Net and C-Net, respectively. The concatenated features are fed
to three prediction heads to estimate the weight vector, the mean vectors, and the covariance matrices of the GMM, respectively. 
 We find empirically that incorporating the historical video frames as the visual context significantly increases computational demands with only slight improvements in performance. Therefore, to keep the scanpath generator lightweight, we choose to omit the visual context. 
The detailed specifications of the density estimation network can be found in the supplementary material. 

Estimating continuous probability density is generally difficult and may lead to overfitting. In particular,  maximum likelihood estimation of the GMM parameters through direct optimization of Eq.~\eqref{eq:GMMcp} is challenging, due to the presence of singularities~\cite{bishop2006pattern}. To circumvent this, we compute the probability mass $P\left(\bar{\boldsymbol r}_t \Big\vert  \boldsymbol{s}_T,\boldsymbol{ c}_{t}\right)$ by discretizing and integrating the density $p\left(\boldsymbol{r}_{t} \Big\vert \boldsymbol{s}_T,\boldsymbol c_{t} \right)$:
\begin{align}
P\left(\bar{\boldsymbol r}_t\vert  \boldsymbol{s}_T,\boldsymbol{c}_{t}\right) =  \int\limits_{\Omega} p\left(\bar{\boldsymbol r}_t \vert  \boldsymbol{s}_T,\boldsymbol{c}_{t} \right) d\Omega.
\end{align}
$\bar{\boldsymbol r}_t$ represents the quantized value of $\boldsymbol r_t$ by a uniform quantizer with a step size of $\Delta$:
\begin{align}\label{eq:quan}
    \bar{\xi} = Q(\xi) = \Delta\left\lfloor\frac{\xi}{\Delta} + \frac{1}{2}\right\rfloor,
\end{align}
\noindent
where $\lfloor\cdot\rfloor$ denotes the floor function. $\Omega =[\bar{\phi}_{T+t}-1/2\Delta,\bar{\phi}_{T+t}+1/2\Delta]\times[\bar{\theta}_{T+t}-1/2\Delta,\bar{\theta}_{T+t}+1/2\Delta]$ is the integration interval. As pointed out  in~\cite{li2023scanpath}, the incorporation of quantization establishes the equivalence between scanpath generation and lossy scanpath compression.

\begin{figure}[!tbp]
\scriptsize
\centering
\vspace{-6pt}
\includegraphics[width=0.4\textwidth]{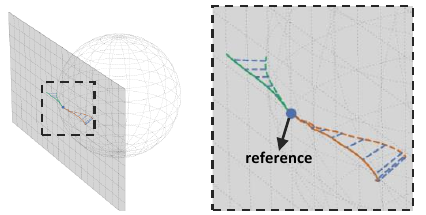}
\caption{Visualization of a relative scanpath projected from the sphere to the viewport.}
\vspace{-11pt}
\label{fig:5}
\end{figure}

Furthermore, the absolute Euler coordinate system is not user-centric, meaning that it is not centered at the user's current viewpoint, relative to historical and future viewpoints~\cite{li2023scanpath}. This may complicate the probabilistic modeling of scanpaths and the end-to-end optimization of blind PVQA. To address this, we convert the Euler coordinates to the relative $uv$ coordinates:
\begin{align}\label{eq:forwardtran}
    \Tilde{\boldsymbol{s}}_{T-t} = \boldsymbol{\Psi}_{T-t}(\boldsymbol{s}_T) ,\, \mathrm{for}\, t\in\{1,\ldots, H\},
\end{align}
\noindent
where $\boldsymbol{\Psi}_{T-t}(\cdot)$ denotes the mapping of $\boldsymbol{s}_T$ to the viewport centered at the reference viewpoint $(\phi_{T-t},\theta_{T-t})$. By choosing each viewpoint in $\boldsymbol{s}_T$ as the reference, we create $H$ relative scanpaths out of $\boldsymbol{s}_T$ (see Figure~\ref{fig:5}), which serve as input to the H-Net. Meanwhile, we map the causal scanpath context $\boldsymbol c_{t}$ and the viewpoint to be predicted $({\phi}_{T+t},{\theta}_{T+t})$ to the last historical viewport centered at $(\phi_{T-1},\theta_{T-1})$. We stack $W$ SGUs to form a scanpath generation block (SGB), which takes  $\boldsymbol{s}_T$ as input and predicts the future scanpath $\boldsymbol{r}$. Furthermore, we stack $M$ SGBs to form the scanpath generator, which is capable of predicting a very long-term scanpath of length $M \times W + H$ (including the initial length $H$). The parameters of different SGUs are shared to enable variable-length scanpath generation by varying $M$.

\noindent\textbf{Differentiable Scanpath Sampling}.
To enable end-to-end optimization of the proposed blind PVQA method, we propose a two-step differentiable sampling method to draw viewpoints from the estimated GMM via the reparameterization trick~\cite{jang2017categorical, kingma2013auto}. The first step is to select a Gaussian component from which to sample the viewpoint, according to  the categorical distribution: 
\begin{align}\label{eq:gmt}
    \boldsymbol{e} = \mathrm{one\_hot}\left(\argmax_{i\in\{1,\ldots K\}}(\mathrm{log}(\alpha_i) + g_i)\right) ,
\end{align}
where $g_i$ is a sample drawn from the $\mathrm{Gumbel}(0,1)$ distribution, and $\boldsymbol{e}$ is a one-hot vector. Eq.~\eqref{eq:gmt} is known as the Gumbel-Max trick~\cite{gumbel1948statistical}, which is non-differentiable. We relax the $\argmax$ operator with a $\mathrm{softmax}$ function~\cite{jang2017categorical}:
\begin{align}\label{eq:gs}
\hat{\boldsymbol{e}} = \mathrm{softmax}((\mathrm{log}(\boldsymbol{\alpha}) + \boldsymbol{g})/\tau),
\end{align}
\noindent
where $\tau$ represents the temperature coefficient, $\boldsymbol{\alpha}=[\alpha_1,\ldots,\alpha_K]$, and $\boldsymbol{g}=[g_1,\ldots,g_K]$. As $\tau$ approaches zero, $\hat{\boldsymbol{e}}$ converges to $\boldsymbol{e}$. In the forward pass, $\argmax$ is used directly, while in the backward pass, it is replaced by the $\mathrm{softmax}$ function
. The second step involves sampling a viewpoint from the selected Gaussian component. Assuming the $i$-th Gaussian component is selected, the linear reparameterization trick~\cite{kingma2013auto} suggests
\begin{align}
    \Tilde{\boldsymbol r}_t =  \boldsymbol{u}_i + \boldsymbol{L}_i \boldsymbol{\epsilon} ,
\end{align}
\noindent
where $\Tilde{\boldsymbol r}_t$ is  the relative $uv$ coordinates of the $t$-th future viewpoint. $\boldsymbol{\Sigma}_i = \boldsymbol{L}_i\boldsymbol{L}_i^\intercal$ is the Cholesky decomposition, and  $\boldsymbol{\epsilon}$ is a sample drawn from the $\mathcal{N}(\mathbf{0}, \mathbf{I})$. Differentiation of the Cholesky decomposition is complicated and sometimes numerically unstable~\cite{smith1995differentiation}. Thus we  assume the independence between the $uv$ coordinates, leading to the simplified reparameterization formula:
\begin{align}
    \Tilde{\boldsymbol{r}}_t =  \boldsymbol{u}_i + \boldsymbol{\sigma}_i \odot \boldsymbol{\epsilon},
\end{align}
where $\odot$ denotes the element-wise product, and $\boldsymbol{\sigma}_i$ denotes the standard deviations of the $i$-th Gaussian component.
Through the two-step reparameterization, our sampling strategy ensures effective back-propagation. As suggested in~\cite{li2023scanpath}, we additionally implement a proportional–integral–derivative (PID)  controller~\cite{ang2005pid} to further improve the smoothness of the sampled scanpaths (see the details in the supplementary material). 

\noindent\textbf{Differentiable Viewport Sequence Generation}. Inspired by~\cite{Jaderberg2015Spatial}, we first parameterize the Euler sampling grid in terms of the relative $uv$ coordinates via the inverse transformation $\boldsymbol{\Psi}^{-1}$ (see Eq.~\eqref{eq:forwardtran}). Subsequently, the Euler coordinate $(\phi, \theta)$  is mapped to the discrete sampling position $(m,n)$ in the ERP domain:
\begin{align}
    m&=(0.5 - \phi/\pi)H_e -0.5 ,\\
    n&=(\theta/2\pi +0.5)W_e-0.5 ,
\end{align}
\noindent
where $H_e$ and $W_e$ are the height and width of the video frame in ERP. Once the mapping between $(u,v)$ and  $(m,n)$ is established, we construct a flow field that is the same size as the viewport. Within this field, each element records the corresponding pixel position $(m,n)$. Given an ERP image and the flow field, we apply bilinear interpolation~\cite{Jaderberg2015Spatial} to compute pixel values in the viewport and leverage its sub-gradients for back-propagation. This process yields $N$ viewport sequences, corresponding to $N$ initial paths.

\subsection{Quality Assessor}
 Our probabilistic scanpath generator can work with any planar VQA model, whether it is differentiable or not. To enable end-to-end optimization of the scanpath generator and the quality assessor, and to make a fair comparison with existing blind PVQA models, we reuse three differentiable quality assessors from ScanpathVQA~\cite{wen2024perceptual},  GSR-S / GSR-X~\cite{sui2023perceptual}, and Assessor360~\cite{wu2023assessor360}. Specifically, the quality assessor of ScanpathVQA is a lightweight ResNet-18 network~\cite{he2016deep}, with the classification head replaced by a quality estimator (a multilayer perceptron). The quality assessors of GSR-S / GSR-X are adapted from Video Swin-T~\cite{liu2022video} / X-Clip-B/32~\cite{ni2022expanding}. The quality assessor of Assessor360 is modified from Swin-B~\cite{liu2021swin} with the addition of a temporal analysis module. We feed each of the $N$ viewport sequences to the quality assessor to compute $N$ quality scores. The overall quality estimate is then computed by a simple  average:
\begin{align}\label{eq:avg}
     \hat{q} = \frac{1}{N}\sum_{i=1}^N \hat{q}_i.
 \end{align}
\subsection{Optimization Strategy}
We explore a three-stage training procedure for our blind PVQA model. In the first stage, we pre-train the density estimation network on the VRVQW dataset~\cite{wen2024perceptual} by minimizing  the expected code length of the generated scanpaths (also equivalent to minimizing the negative log-likelihood):
\begin{align}
\ell_\mathrm{code} =  -\frac{1}{B W}\sum_{i=1}^{B} \sum_{t=0}^{W-1} \log _2 \left(P\left(\bar {\boldsymbol r}_t^{(i)} \Big\vert  \boldsymbol{s}^{(i)},\boldsymbol{ c}^{(i)}_{t}\right)\right) ,
\end{align}
\noindent
where $B$ denotes the mini-batch size. During this stage of training, we use human scanpaths to fill in the causal path context, which can be efficiently implemented by a causal masking mechanism.  In the second stage, we fix the parameters of the pre-trained scanpath generator and warm up the quality assessor by optimizing the Pearson linear correlation coefficient (PLCC) between human perceptual scores and model predictions. In the third stage, we end-to-end finetune the entire method. We find that the proposed three-stage optimization strategy accelerates convergence compared to the naive end-to-end optimization.

\section{Experiments}
In this section, we first delineate the experimental setups, and then compare our method with current blind PVQA models under both in-dataset and cross-dataset settings. We further validate our scanpath generator in terms of explaining human perceptual scores and replicating human viewing patterns. Lastly, we conduct a series of ablation experiments to probe the impact of several key designs.

\subsection{Experimental Setups}

\noindent\textbf{Datasets}. We employ three panoramic image and video datasets: VRVQW~\cite{wen2024perceptual}, CVIQD~\cite{sun2018cviqd}, and OIQA~\cite{duan2018perceptual}. The VRVQW dataset includes $502$ \textit{panoramic videos} that have a wide spectrum of authentic distortions. Each video is viewed under four unique viewing conditions to simulate the different quality of experience during the initial viewing. The CVIQD dataset comprises a total of $528$ compressed \textit{panoramic images} by JPEG, AVC, and HEVC, from $16$ reference images. The OIQA dataset includes $320$ \textit{panoramic images} that have been altered from $16$ reference images by JPEG compression, JPEG2000 compression, Gaussian blurring, and Gaussian noise contamination.

\begin{table}
 \centering 
 \footnotesize 
 \caption{In-dataset comparison of blind PVQA methods on three panoramic image and video quality datasets. 
 SRCC: Spearman's rank correlation coefficient. PLCC: Pearson linear correlation coefficient. The same evaluation metrics are applied in Tables~\ref{tab:3},~\ref{tab:4},~\ref{tab:6},~\ref{tab:7} and Figure~\ref{fig:3}. The best results on each dataset are highlighted in bold. }.
 \vspace{-6pt}
 \begin{tabular}{l  l  c c}
     \toprule 
     Dataset & Method & SRCC & PLCC\\
     \midrule
     &NIQE~\cite{mittal2012making} & $0.401$ & $0.365$  \\
     &MC360IQA~\cite{Sun2020MC360IQA}  & $0.669$ & $0.671$ \\
     &Wen24~\cite{wen2024perceptual} & $0.756$ & $0.763$ \\ 
     & ScanpathVQA~\cite{wen2024perceptual}  & $0.779$ & $0.781$ \\
      VRVQW~\cite{wen2024perceptual}  & Assessor360~\cite{wu2023assessor360}  &$0.406$&$0.415$ \\
      \cmidrule{2-4}
     & Ours (ScanpathVQA) & $0.801$ & $0.809$ \\
     & Ours (GSR-S) & $0.804$ & $0.807$ \\
     & Ours (GSR-X) & $0.815$ & $0.819$ \\
     & Ours (Assessor360) & $\mathbf{0.822}$ & $\mathbf{0.823}$ \\
     \midrule
     &NIQE~\cite{mittal2012making} & $0.847$ & $0.878$  \\
     &MC360IQA~\cite{Sun2020MC360IQA} & $0.917$ & $0.939$ \\
     &Wen24~\cite{wen2024perceptual} & $0.919$ & $0.932$ \\ 
     & GSR-S~\cite{sui2023perceptual} & $0.905$ & $0.937$ \\
     & GSR-X~\cite{sui2023perceptual} & $0.944$ & $0.962$  \\
     CVIQD~\cite{sun2018cviqd}& Assessor360~\cite{wu2023assessor360} & $0.955$ & $0.969$  \\
     \cmidrule{2-4}
     & Ours (ScanpathVQA) & $0.912$ & $0.936$ \\
     & Ours (GSR-S) & $0.930$ & $0.958$ \\
     & Ours (GSR-X) & $0.956$ & $0.974$ \\
     & Ours (Assessor360) & $\mathbf{0.972}$ & $\mathbf{0.983}$ \\
     \midrule
     &NIQE~\cite{mittal2012making} & $0.702$ & $0.657$ \\
     &MC360IQA~\cite{Sun2020MC360IQA} & $0.900$ & $0.906$ \\
     &Wen24~\cite{wen2024perceptual} & $0.905$ & $0.907$ \\ 
     &GSR-S~\cite{sui2023perceptual} & $0.902$ & $0.915$  \\
     &GSR-X~\cite{sui2023perceptual} & $0.945$ & $0.954$ \\
     OIQA~\cite{duan2018perceptual}&Assessor360~\cite{wu2023assessor360} & $0.946$ & $0.953$ \\
     \cmidrule{2-4}
     &Ours (ScanpathVQA) & $0.915$ & $0.922$ \\
     &Ours (GSR-S) & $0.927$ & $0.936$ \\
     &Ours (GSR-X) & $0.956$ & $0.967$ \\
     &Ours (Assessor360) & $\mathbf{0.960}$ & $\mathbf{0.971}$ \\
     \bottomrule
    \end{tabular}
\label{tab:1}
\end{table}

\begin{table}[t]
 \centering 
 \footnotesize
 \caption{Cross-dataset comparison of blind PVQA methods on the CVIQD~\cite{sun2018cviqd} and OIQA~\cite{duan2018perceptual} datasets. The arrow points from the training set to the test set.}  
\vspace{-6pt}
 \begin{tabular}{c  l  c  c}
     \toprule 
     Dataset & Method & SRCC & PLCC  \\
     \midrule
     &MC360IQA & $0.798$ & $0.842$ \\ 
     OIQA&GSR-X & $0.762$ & $0.841$ \\
     $\downarrow$&
    Assessor360 & $0.859$ & $0.893$ \\ 
    \cmidrule{2-4}
     CVIQD&
     Ours (ScanpathVQA) & $0.733$ & $0.747$ \\  
     &Ours (Assessor360) & $\mathbf{0.872}$ & $\mathbf{0.904}$\\
     \midrule 
     &MC360IQA & $0.288$ & $0.349$ \\ 
     CVIQD &GSR-X & $0.695$ & $\mathbf{0.718}$ \\ 
     $\downarrow$
      &Assessor360 & $0.338$ & $0.467$ \\
      \cmidrule{2-4}
     OIQA
     &Ours (ScanpathVQA) & $0.636$ & $0.658$ \\
     &Ours (Assessor360) & $\mathbf{0.703}$ & $0.715$ \\
     \bottomrule
    \end{tabular}
\vspace{-11pt}
\label{tab:3}
\end{table}

\noindent\textbf{Implementation Details}.
For the scanpath generator, we set the length of the provided initial path $H$ and the predicted future path $W$ in the SGB to be identical and equal to $5$. The number of Gaussian components $K$ in Eq.~\eqref{eq:GMMcp} is set to $3$. The
quantization step size $\Delta$ in Eq.~\eqref{eq:quan} is set to $0.2$. 
The temperature coefficient $\tau$ in Eq.~\eqref{eq:gs} is set to $1$. 
The number of stacked SGBs $M$ is set to $6$ and $14$ for the viewing duration of $7$ and $15$ seconds, respectively. 
For the quality assessor,
the number of scanpaths $N$ in Eq.~\eqref{eq:avg} is set to $20$. The input viewport size $H_v \times W_v$ is $224 \times 224$, corresponding to a field of view of $90^\circ \times 90^\circ$. 
The length of the viewport sequence is set to $7$ regardless of the duration and frame rate of the original panoramic video. We split VRVQW randomly into the training, validation, and test sets according to the ratio of $6:2:2$ for $5$ times, and report the mean results. Similarly,
 we split CVIQD and OIQA using a different ratio of $7:1:2$ for $5$ times. The detailed configuration of our three-stage optimization strategy is detailed in the supplementary material.

\begin{table}[t]
 \centering 
 \footnotesize
 \caption{Comparison of different scanpath generators for explaining human perceptual scores.}  
 \vspace{-6pt}
 \begin{tabular}{l  l  c  c}
     \toprule
      Dataset & Method & SRCC & PLCC \\
     \midrule
     & \textit{Human scanpath} & $0.786$ & $0.790$ \\ 
     & Random sampling & $0.075$ & $0.104$ \\ 
     
     & Heuristic sampling~\cite{wu2023assessor360} & $0.431$ & $0.443$ \\ 
     VRVQW &Xu18~\cite{xu2018gaze} & $0.712$ & $0.717$ \\
     &TRACK~\cite{rondon2022track} & $0.745$  &  $0.749$ \\
     & Li23~\cite{li2023scanpath} & $0.790$ & $0.794$ \\ 
     & Ours & $\mathbf{0.805}$ & $\mathbf{0.814}$ \\
     \midrule
      &Random sampling & $0.632$ & $0.640$ \\ 
      
      & Heuristic sampling~\cite{wu2023assessor360} & $0.868$ & $0.870$ \\ CVIQD&ScanDMM~\cite{sui2023scandmm} & $0.856$ & $0.864$ \\
      &Li23~\cite{li2023scanpath} & $0.814$ & $0.827$ \\
      &Ours & $\mathbf{0.928}$ & $\mathbf{0.940}$\\
     \midrule
     &Random sampling & $0.514$ & $0.536$ \\ 
     &Heuristic sampling~\cite{wu2023assessor360} & $0.861$ & $0.872$ \\
     OIQA
     &ScanDMM~\cite{sui2023scandmm} & $0.865$ & $0.877$ \\ 
     &Li23~\cite{li2023scanpath} & $0.793$ & $0.799$ \\
     &Ours & $\mathbf{0.914}$ & $\mathbf{0.917}$ \\
     \bottomrule
    \end{tabular}
\vspace{-11pt}
\label{tab:4}
\end{table}

\subsection{Main Results}
We compare our blind PVQA method with seven existing models, including NIQE~\cite{mittal2012making}, MC360IQA~\cite{Sun2020MC360IQA}, Wen24~\cite{wen2024perceptual}, ScanpathVQA~\cite{wen2024perceptual}, GSR-S~\cite{sui2023perceptual}, GSR-X~\cite{sui2023perceptual}, and Assessor360~\cite{wu2023assessor360}. For image quality models such as MC360IQA~\cite{Sun2020MC360IQA} and Assessor360~\cite{wu2023assessor360}, we retrain them on the VRVQW dataset. In the case of MC360IQA, we assign the video-level quality score to each key frame and use their temporally averaged score for testing. Assessor360's scanpath generator is adapted to videos by adjusting the semantic context associated with each key frame.

\noindent\textbf{In-dataset Results}. Table~\ref{tab:1} shows the Spearman's rank correlation coefficient
(SRCC) and PLCC\footnote{As standard practice, we apply a monotonic logistic function to compensate for the nonlinearity in model predictions before computing PLCC.} results under the in-dataset setting. It is evident that our learned scanpaths enhance the performance of all quality assessors compared to other scanpath-based methods. When integrated with the quality assessor from Assessor360 (\ie, a modified Swin-B with a temporal analysis module), our proposed method achieves the best results on all three datasets. Furthermore, our scanpath generator can boost a simpler quality assessor (\eg, from ScanpathVQA~\cite{wen2024perceptual} with approximately $11$ million parameters) to reach performance levels similar to those of a more sophisticated quality assessor coupled with a weaker scanpath generator (\eg, GSR-S~\cite{sui2023perceptual} with approximately $112$ million parameters). Methods that overlook human viewing patterns, like NIQE~\cite{mittal2012making} and MC360IQA~\cite{Sun2020MC360IQA}, fail to accurately model the human perception of panoramic image and video quality, especially on VRVQW.  Additionally, we find that assessing the quality of panoramic videos with authentic distortions tends to be more challenging than for panoramic images with synthetic distortions. This is anticipated because authentic distortions in panoramic videos typically present as a complex blend of various artifacts, localized in space and time.

\noindent\textbf{Cross-dataset Results}. Table~\ref{tab:3} shows the SRCC and PLCC results under the cross-dataset settings on CVIQD~\cite{sun2018cviqd} and OIQA~\cite{duan2018perceptual}. Generally,
models trained on the OIQA dataset show better generalizability. This is likely because OIQA encompasses
a broader range of distortion types, compared to CVIQD, which only includes the compression artifacts. Assessor360 shows a noticeable performance drop when tested on OIQA, potentially indicative of overfitting. 
 Our scanpath generator is capable of restoring 
 Assessor360's performance, which provides a strong indication of its effectiveness through end-to-end optimization.
 

\begin{table}[tb]
 \centering 
 \footnotesize
 \caption{Comparison of different scanpath generators for replicating human viewing patterns using the minimum orthodromic distance (minOD) and maximum temporal correlation (maxTC).}  
 \vspace{-6pt}
 \begin{tabular}{l  c  c}
     \toprule 
      Method & $\mathrm{minOD}\downarrow$ & $\mathrm{maxTC}\uparrow$  \\
     \midrule
      Heuristic sampling~\cite{wu2023assessor360} & $1.325$ & $0.401$ \\
     Xu18~\cite{xu2018gaze} & $1.185$ & $0.637$ \\
     TRACK~\cite{rondon2022track} & $1.067$ & $0.699$ \\
     Li23~\cite{li2023scanpath} & $0.542$ & $0.796$ \\
     \midrule
     Ours (w/o end-to-end optimization)& $0.556$ & $0.781$ \\
     Ours & $\mathbf{0.536}$ & $\mathbf{0.805}$ \\
     \bottomrule
    \end{tabular}
\label{tab:8}
\end{table}

\begin{figure}[!tbp]
\scriptsize
\centering
\includegraphics[width=0.72\linewidth]{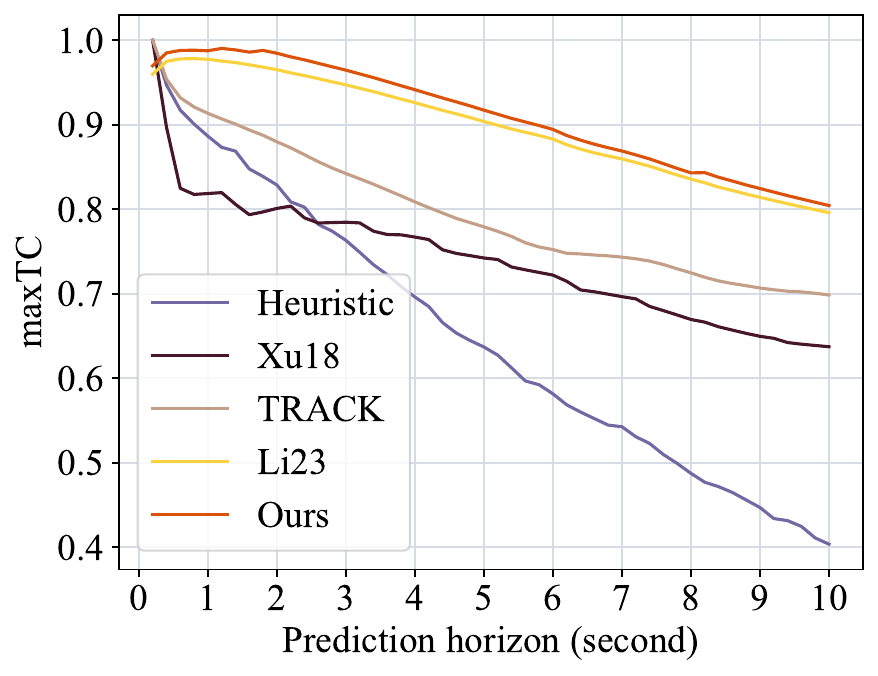}
\vspace{-6pt}
\caption{Comparison of different scanpath predictors in terms of $\mathrm{maxTC}$ with different prediction horizons.}
\vspace{-11pt}
\label{fig:4}
\end{figure}

\subsection{Scanpath Generator Validation}
\noindent\textbf{Explaining Human Perceptual Scores}. We conduct an apple-to-apple comparison of different scanpath generators in terms of explaining human perceptual scores by fixing the quality assessor to that used in ScanpathVQA.
These include random sampling, heuristic sampling~\cite{wu2023assessor360}, Xu18~\cite{xu2018gaze}, TRACK~\cite{rondon2022track}, ScanDMM~\cite{sui2023scandmm}, Li23~\cite{li2023scanpath}, and our method. 
Table~\ref{tab:4} shows the SRCC and PLCC results 
on the VRVQW, CVIQD and OIQA datasets. It is noteworthy that our scanpath generator outperforms all competing methods across all three datasets, even surpassing the \textit{human-level} performance on VRVQW. The performance of random and heuristic sampling decreases sharply on VRVQW, due to the presence of spatiotemporally localized authentic distortions. 

\noindent\textbf{Replicating Human Viewing Patterns}. We also test different scanpath generators in terms of replicating human viewing patterns on the VRVQW dataset by comparing the predicted scanpaths to those of humans. We use two set-to-set evaluation metrics: the minimum orthodromic distance (\ie, minOD) and maximum temporal correlation (\ie, maxTC), as suggested in~\cite{li2023scanpath}. Given a set of human scanpaths, $\mathcal{S} = \{\boldsymbol s^{(i)}\}_{i=1}^{\vert\mathcal{S}\vert}$, the minimum orthodromic distance between $\mathcal{S}$ and the set of predicted scanpaths $\hat{\mathcal{S}}=\{\hat{\boldsymbol s}^{(i)}\}_{i=1}^{\vert
\hat{\mathcal{S}}\vert}$ can be computed by
\begin{align}
\begin{split} 
\mathrm{minOD}\left(\mathcal{S},\hat{\mathcal{S}}\right)=\min\limits_{\boldsymbol s\in \mathcal{S}, \hat{\boldsymbol s}\in \hat{\mathcal{S}}}\mathrm{OD}\left(\boldsymbol s,\hat{\boldsymbol s}\right),
\end{split}
\end{align}
where the orthodromic distance $\mathrm{OD}(\cdot,\cdot)$ is defined as 
\begin{align}
\begin{split}
    \mathrm{OD}(\boldsymbol s,\hat{\boldsymbol s}) = \frac{1}{T}\sum_{t=0}^{T-1}&\mathrm{arccos}\Big(\mathrm{cos}(\phi_{t})\mathrm{cos}(\hat{\phi}_{t})\mathrm{cos}(\theta_t - \hat{\theta}_t)
    \\&+\mathrm{sin}(\phi_{t})\mathrm{sin}(\hat{\phi}_{t})\Big).
\end{split}
\end{align}
The maximum temporal correlation between $\mathcal{S}$ and $\hat{\mathcal{S}}$ is defined as 
\begin{align}
\mathrm{maxTC}\left(\mathcal{S}, \hat{\mathcal{S}}\right)=\max\limits_{\boldsymbol s\in \mathcal{S}, \hat{\boldsymbol s}\in \hat{\mathcal{S}}}\mathrm{TC}(\boldsymbol{s},\hat{\boldsymbol s}),
\end{align}
where the temporal correlation is computed by
\begin{align}
\begin{split}
    &\mathrm{TC}\left(\boldsymbol s^{(i)},\boldsymbol s^{(j)}\right)=\\&
    \frac{1}{2}\left(\mathrm{PLCC}\left(\boldsymbol \phi^{(i)},\boldsymbol \phi^{(j)}\right) +\mathrm{PLCC}\left(\boldsymbol \theta^{(i)},\boldsymbol \theta^{(j)}\right)\right).
\end{split}
\end{align}

\begin{figure}[!tbp]
  \centering
  \includegraphics[width=0.99\linewidth]{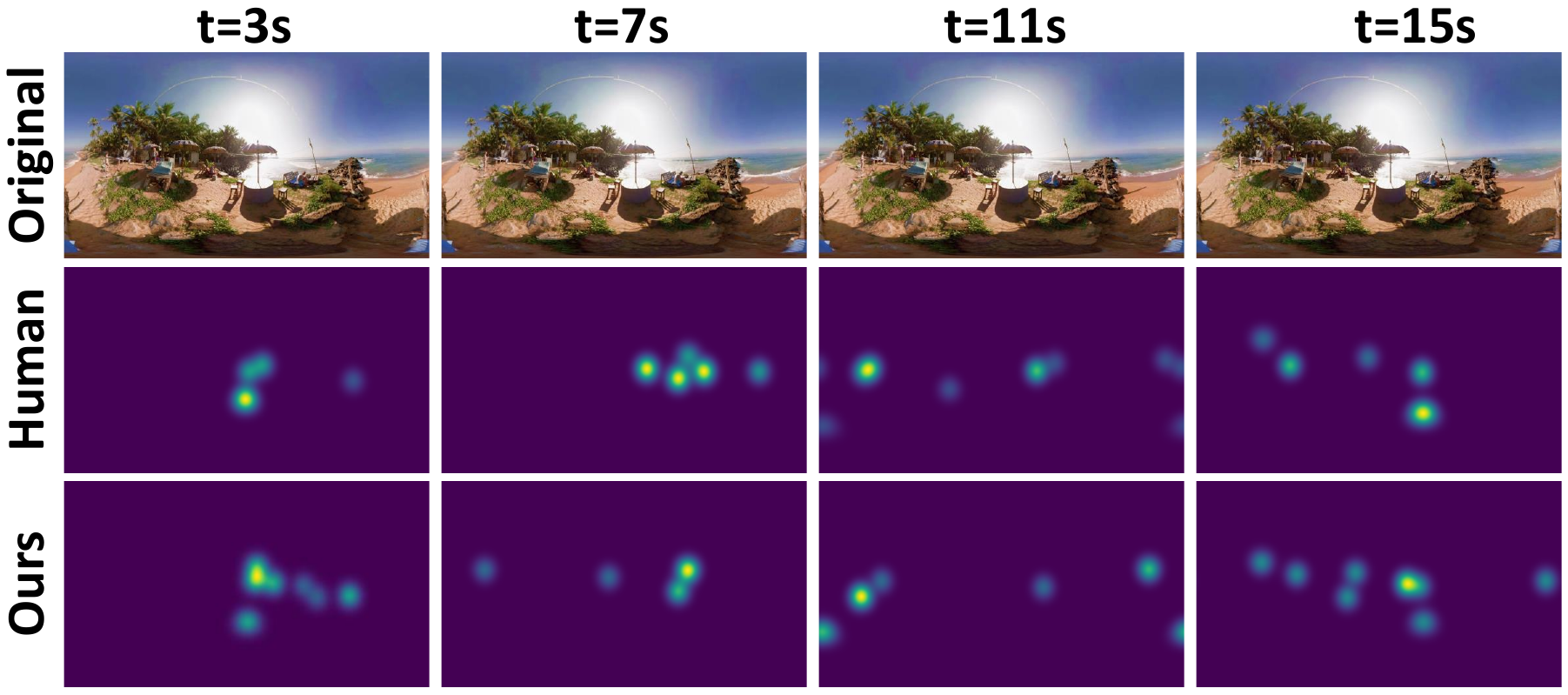}
\vspace{-6pt}
  \caption{Comparison of saliency maps generated from scanpaths by our method and those by humans.}
  \label{fig:r1}
\end{figure}

\begin{table}\footnotesize
\centering
\caption{Impact of optimization strategies on blind PVQA. Training protocol: 1) Two-stage w/ fixed scanpaths, 2) Two-stage w/ varied scanpaths, and 3) Three-stage w/ end-to-end optimization.}
\vspace{-6pt}
\label{tab:6}
\begin{tabular}{c|cc|cc|cc}
\toprule
\multirow{2}{*}{Protocol}  & \multicolumn{2}{c|}{VRVQW} & \multicolumn{2}{c|}{CVIQD} & \multicolumn{2}{c}{OIQA}\\
  &   SRCC  & PLCC & SRCC & PLCC & SRCC & PLCC \\
\midrule
 1& $0.769$ & $0.772$ & $0.710$ & $0.780$ & $0.688$ & $0.742$ \\ 
 2 & $0.781$ & $0.785$ & $0.830$ & $0.859$ & $0.798$ & $0.815$\\ 
 3 & $\mathbf{0.805}$ & $\mathbf{0.814}$ & $\mathbf{0.928}$ & $\mathbf{0.940}$ & $\mathbf{0.914}$ & $\mathbf{0.917}$\\ 

\bottomrule
\end{tabular}
\vspace{-11pt}
\end{table}

\begin{table*}\footnotesize
\centering
\caption{Impact of visual context on blind PVQA. \#Parameters added to the scanpath generator are also shown.}
\vspace{-6pt}
\label{tab:7}
\begin{tabular}{l|c|cc|cc|cc}
\toprule
\multirow{2}{*}{Method} & \multirow{2}{*}{\#Parameters} & \multicolumn{2}{c|}{VRVQW} & \multicolumn{2}{c|}{CVIQD} & \multicolumn{2}{c}{OIQA}\\
  & &  SRCC  & PLCC & SRCC & PLCC & SRCC & PLCC \\
\midrule
 Ours & $1$M & $0.805$ & $0.814$ & $0.928$ & $0.940$ & $0.914$ & $0.917$ \\ 
 Ours w/ visual context & $27$M & $\mathbf{0.816}$ & $\mathbf{0.823}$ & $\mathbf{0.937}$ & $\mathbf{0.956}$ & $\mathbf{0.925}$ & $\mathbf{0.934}$\\ 
\bottomrule
\end{tabular}
\vspace{-6pt}
\end{table*}

Table~\ref{tab:8} presents the minOD and maxTC results, from which we find that our end-to-end optimized scanpath generator delivers the best results, surpassing its independently optimized counterpart by a clear margin. The heuristic sampling~\cite{wu2023assessor360} that depends on the simplified entropy features and equator bias, is inadequate for capturing human viewing patterns, especially for long-term prediction horizons (see Figure~\ref{fig:4}). Due to the deterministic nature,  Xu18~\cite{xu2018gaze} and TRACK~\cite{rondon2022track} fail to accommodate the
diversity and uncertainty inherent in human scanpaths, resulting in subpar performance. Incorporating historical video frames as the visual context, Li23~\cite{li2023scanpath} shows performance on par with our method, reinforcing our assertion that visual context informs less about future viewpoints. Figure~\ref{fig:r1} demonstrates a comparison of the saliency maps derived from scanpaths by our method and those by humans, offering further proof of the close alignment of our scanpath generator and human viewing behaviors.

\subsection{Ablation Studies}

\noindent\textbf{Impact of Optimization Strategies}.
We explore three different optimization strategies: 1) a two-stage approach where the pre-trained scanpath generator produces
a fixed set of scanpaths for the training of the quality assessor, 2) a similar two-stage approach 
but supplying a varied set of scanpaths in each epoch of training, and 3) the default three-stage approach that enables end-to-end optimization. From the results in Table~\ref{tab:6}, we find that the two-stage approach benefits from ``data augmentation'' with varied scanpaths in each epoch.  Our three-stage end-to-end optimization strategy further boosts the accuracy of quality prediction by jointly finetuning both the scanpath generator and quality assessor. 

\noindent\textbf{Impact of Visual Context}.
To assess the impact of visual context on blind PVQA, we enhance our scanpath generator with a video analysis network~\cite{li2023scanpath}, 
 implemented by a variant of ResNet-50 for frame-level feature extraction and aggregation. We 
subsequently integrate it with the ScanpathVQA quality assessor for blind PVQA, with the results shown in Table~\ref{tab:7}. We find that the visual context has a negligible effect on blind PVQA, and thus we exclude it in the generation of scanpaths.

\begin{figure}[!tbp]
\scriptsize
\centering
\scriptsize
\centering
\includegraphics[width=0.72\linewidth]{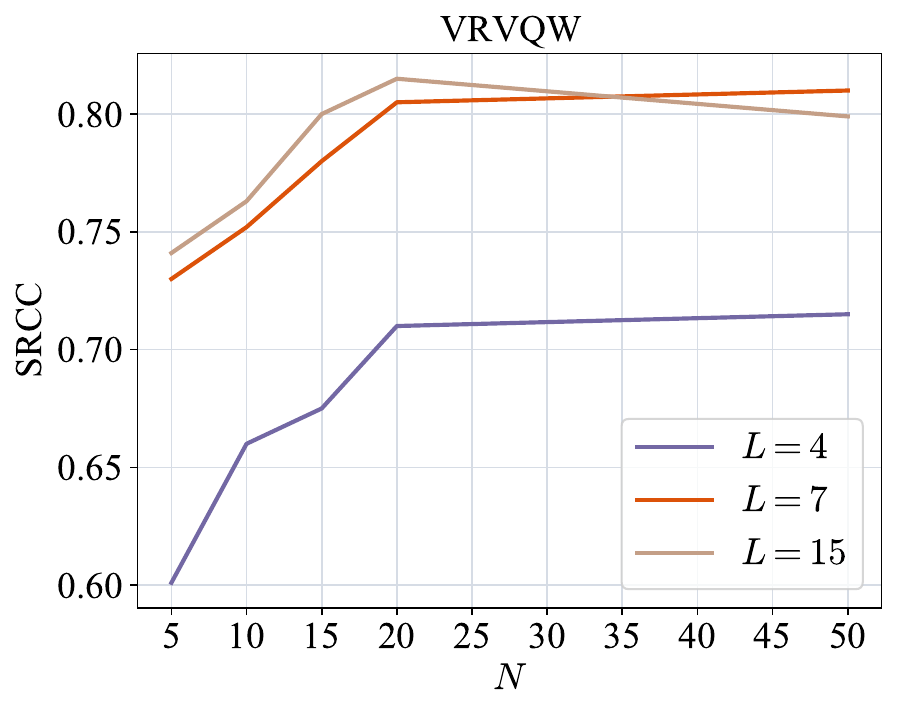}
\vspace{-6pt}
\caption{ 
Impact of the number $N$ and length $L$ of viewport sequences on blind PVQA.}
\vspace{-11pt}
\label{fig:3}
\end{figure}

\noindent\textbf{Impact of the  Number and Length of Viewport Sequences}.
 We investigate the effects of varying the number $N$ and length $L$ of viewport sequences on blind PVQA.  We test $N$ values from $\{5,10,15,20,50\}$ and $L$ values from $\{4,7, 15\}$. Figure~\ref{fig:3} shows the SRCC results on VRVQW, using the ScanpathVQA quality assessor for prediction.  It is clear that $N=20$ viewport sequences are sufficient for reliable quality assessment, with performance remaining stable with the increase in $N$.  For sequence length, $L=7$ appears to be a wise choice. Further increasing $L$ does not noticeably affect the 
 performance, but would lead to a considerable rise in computational demand. Conversely, a shorter viewport sequence results in a noticeable drop in performance due to the loss of information from excessive temporal downsampling. It is important to note that these findings are specific to the ScanpathVQA quality assessor and may differ from other assessors.
 
\section{Conclusion}
We have introduced an end-to-end optimized blind PVQA method, consisting of a scanpath generator and a quality assessor. The proposed scanpath generator is differentiable and can be integrated with any planar VQA model, whose effectiveness has been thoroughly validated in supporting blind PVQA and in modeling human viewing patterns. Additionally, we have also devised a three-stage optimization strategy to facilitate training convergence, which aligns with current large-scale optimization practices that
involve self-supervised pre-training followed by supervised finetuning, including initial warmup phases~\cite{he2022masked}. 

\section{Acknowledgement}

This work was supported in part by the National Natural Science Foundation of China (62071407 
 and 62102339), the Shenzhen Science and Technology Program (RCBS20221008093121052), the Research Grants Council of Hong Kong (ECS 2121382, ECS 27212822, and GRF 17201822), and the CCF-Tencent Rhino Bird Fund (9239061).

{
    \small
    \bibliographystyle{ieeenat_fullname}
    \bibliography{main}
}

\clearpage
\maketitlesupplementary
\paragraph{\fontsize{12}{12}\selectfont{1. Details of the Density Estimation Network}}\mbox{}\\
\vspace{-4pt}

\begin{figure*}[hbt!]
\scriptsize
\centering
\includegraphics[width=1\textwidth]{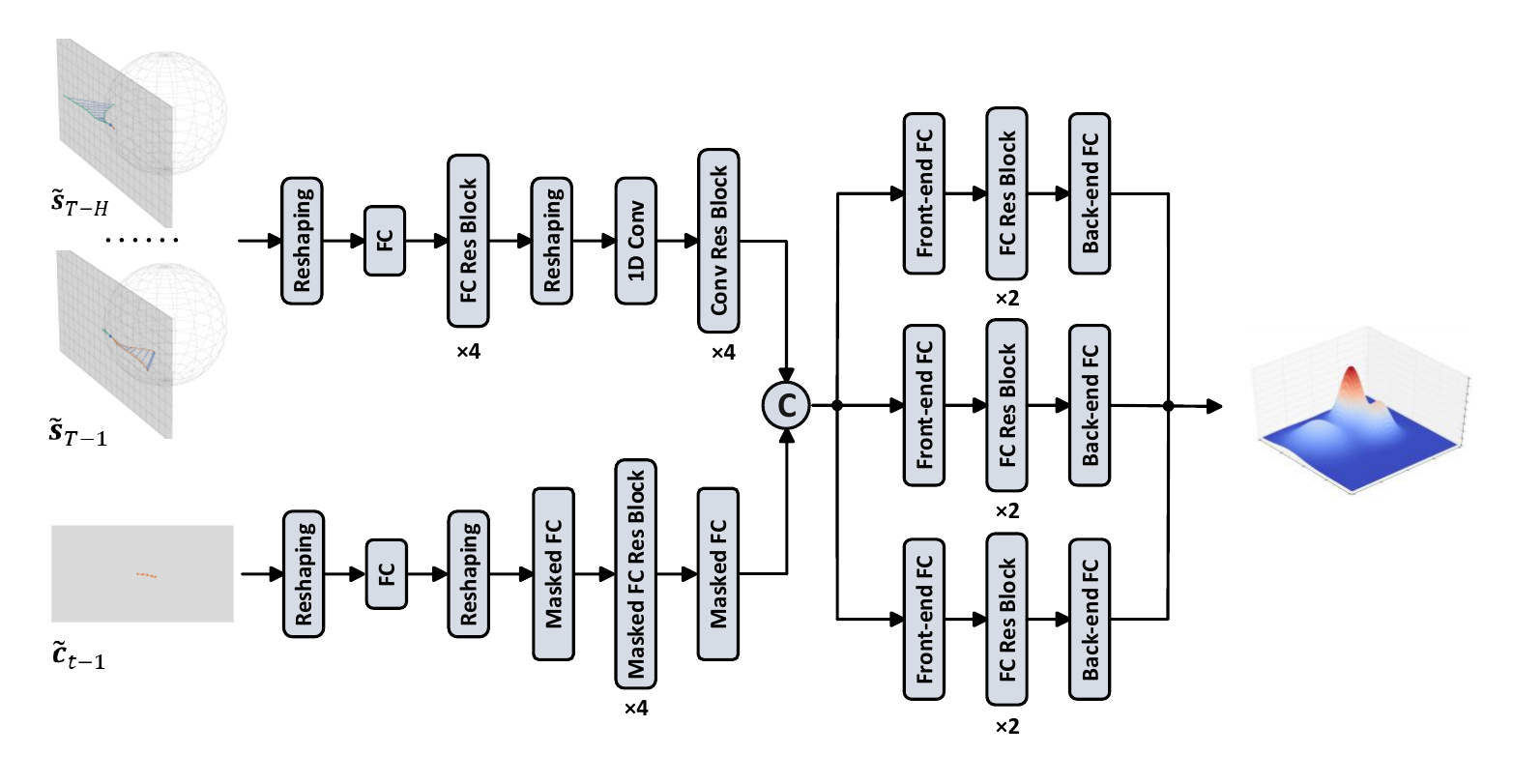}
\vspace{-12pt}
\caption{Architecture of the density estimation network.}
\vspace{-9pt}
\label{fig:7}
\end{figure*}

\noindent 
The architecture of the density estimation network is depicted in Figure~\ref{fig:7}. Of particular interest is the masked computation~\cite{li2023scanpath} used to process the causal path. We refer the readers to the code implementation at \url{https://github.com/kalofan/AutoScanpathQA} for detailed parameter configurations.

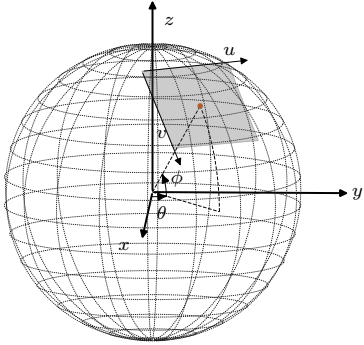
\begin{figure}[!tbp]
\centering
\begin{subfigure}{0.6\linewidth}
\begin{tikzpicture}
\node (nd1) at (0,1.3) {\includegraphics[width=0.95\linewidth]{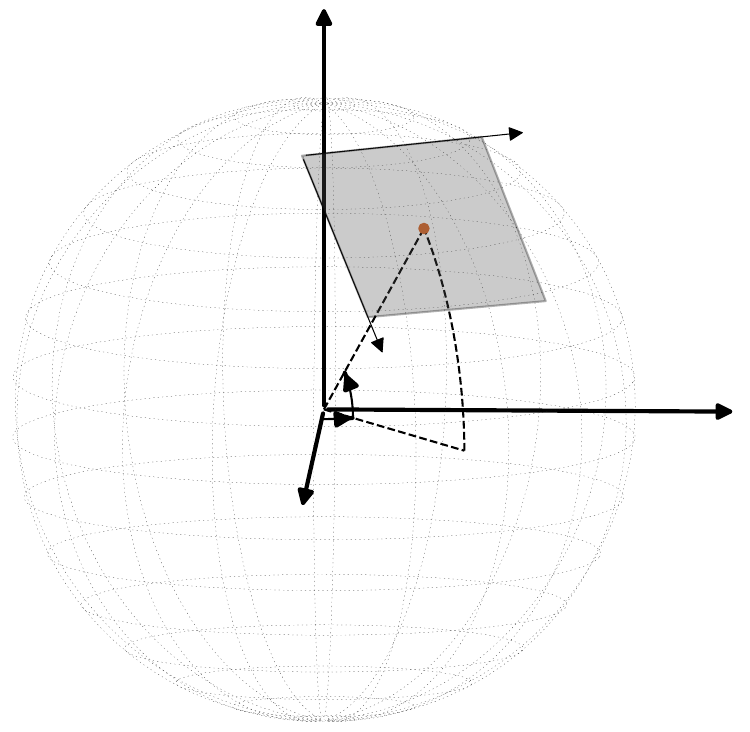}};
\node[text width=0.5cm,align=center] (t1) at (0,1.2) {\baselineskip=3pt \scriptsize{$\phi$} \par};
\node[text width=0.5cm,align=center] (t2) at (-0.2,0.75) {\baselineskip=3pt \scriptsize{$\theta$} \par};
\node[text width=0.5cm,align=center] (t3) at (2.4,1) {\baselineskip=3pt \scriptsize{$y$} \par};
\node[text width=0.5cm,align=center] (t3) at (-0.1,3.3) {\baselineskip=3pt \scriptsize{$z$} \par};
\node[text width=0.5cm,align=center] (t3) at (-0.7,0.3) {\baselineskip=3pt \scriptsize{$x$} \par};
\node[text width=0.5cm,align=center] (t3) at (0.7,2.9) {\baselineskip=3pt \scriptsize{$u$} \par};
\node[text width=0.5cm,align=center] (t2) at (-0.2,1.8) {\baselineskip=3pt \scriptsize{$v$} \par};
\end{tikzpicture}
\end{subfigure}
\caption{Illustration of different coordinate systems relevant to panoramic signal processing:  Spherical Euler coordinates $(\phi, \theta)$, 3D Euclidean coordinates $(x, y, z)$, and $uv$ coordinates $(u,v)$. Image adapted from~\cite{li2023scanpath}.}
\label{fig:8}
\end{figure}

\paragraph{\fontsize{12}{12}\selectfont{2. Relative $uv$ Coordinate System}}\mbox{}\\
\vspace{-4pt}

\noindent
Figure~\ref{fig:8} compares different coordinate systems. The transformation $(u, v) = \boldsymbol{\mathrm{\Psi}}_t(\phi,\theta)$ that maps the viewpoint $(\phi,\theta)$ to the $(u,v)$ coordinates relative to the $t$-th viewport centered at $(\phi_t,\theta_t)$ is broken down into multiple steps. First, $(\phi,\theta)$ is  transformed to $(x,y,z)$ in the Cartesian coordinate system:
\begin{align}
 x&=r\cos(\phi)\cos(\theta), \\
 y&=r\cos(\phi)\sin(\theta),  \\
 z&=r\sin(\phi), 
\end{align}
where $r=0.5 W_v \cot(0.5 \theta_v)$ is the radius of the sphere, determined by the width of the viewport, $W_v$ and the field of view, $\theta_v$. 
Second, $(x,y,z)$ is rotated with respect to $(\phi_t, \theta_t)$:
\begin{align}
\begin{pmatrix}
x_t\\
y_t\\
z_t
\end{pmatrix}
= R(\phi_t, \theta_t)^\intercal  \times 
\begin{pmatrix}
x \\
y \\
z
\end{pmatrix}.
\end{align}
where $R(\phi_t, \theta_t)$ is the rotation matrix defined as the product of two matrices $R_2\times R_1$:
\begin{align}
R_1 &= \begin{pmatrix}
a & -b & 0 \\
b & a & 0 \\
0 & 0 & 1
\end{pmatrix}, &\\
R_2 &= \begin{pmatrix}
c + (1-c) b^2 & -(1-c) a b & -d a \\
-(1-c) a b & c + (1-c) a^2 & -d b \\
d a & d b & c
\end{pmatrix},
\end{align}
where 
\begin{align}
    a&=\cos(\theta_t),\\
    b&=\sin(\theta_t),\\
    c&=\cos(\phi_t),\\
    d&=\sin(\phi_t).
\end{align}
After rotation, $(x_t,y_t,z_t)$ is transformed to $(r,y_t^\prime,z_t^\prime)$ by projecting it to the plane $x=r$. Last, $(r,y_t^{\prime},z_t^{\prime})$ is readily represented in the $uv$ coordinate system:
\begin{align}
u&=y_t^\prime + 0.5W_v - 0.5 ,\\
v&=0.5H_v - z_t^\prime - 0.5 ,
\end{align}
where $H_v$ is the height of the viewport. We may shift the origin to the viewport center, leading to
\begin{align}
u&= y_t^{\prime}, \\
v&= -z_t^{\prime} .
\end{align}

\paragraph{\fontsize{12}{12}\selectfont{3. PID Controller}}\mbox{}\\
\vspace{-4pt}

\noindent
The PID controller~\cite{ang2005pid} is a prevalent feedback mechanism that enables continuous modulation of control signals. We adopt the sampling strategy of~\cite{li2023scanpath} and assume a proxy viewer governed by Newton's laws of motion. Initially, the proxy viewer is positioned at the starting point \(\hat{\boldsymbol r}_{-1}=(0,0)\) in the \(uv\) coordinate system, with an initial speed \(\boldsymbol b_{-1}\) and acceleration \(\boldsymbol a_{-1}\). We determine the \(t\)-th  viewpoint by
\begin{align}\label{eq:nm}
    \hat{\boldsymbol r}_t = \hat{\boldsymbol r}_{t-1} + \Delta\lambda\boldsymbol b_{t-1} + \frac{1}{2}(\Delta\lambda)^2\boldsymbol a_{t-1},  t\in \{0,\ldots, W-1\},
\end{align}
\noindent
where the speed \(\boldsymbol b_{t-1}\) is updated by
\begin{align}
    \boldsymbol b_t = \boldsymbol b_{t-1} + \Delta\lambda\boldsymbol a_{t-1}.
\end{align}
\noindent
\(\Delta\lambda\) represents the sampling interval, \ie, the inverse of the sampling rate. To update the acceleration \(\boldsymbol a_{t-1}\), a reference viewpoint \(\Tilde{\boldsymbol r}_t\) for \(\hat{\boldsymbol r}_t\) is provided by sampling from \(P\left(\Tilde{\boldsymbol r}_t\Big\vert \boldsymbol{s}, \boldsymbol{c}_t\right)\), where \(\boldsymbol c_t =\{\hat{\boldsymbol r}_0, \ldots, \hat{\boldsymbol r}_{t-1}\}\). Consequently, an error signal is generated:
\begin{align}
    \boldsymbol e_t = \Tilde{\boldsymbol r}_t - \hat{\boldsymbol r}_t,
\end{align}
\noindent
which is fed to the PID controller for acceleration adjustment:
\begin{align}\label{eq:pida}
    \boldsymbol a_t = K_p \boldsymbol e_t + K_i \sum_{\tau=0}^{t}\boldsymbol e_\lambda + K_d (\boldsymbol e_t - \boldsymbol e_{t-1}),
\end{align}
\noindent
where \(K_p\), \(K_i\), and \(K_d\) denote the proportional, integral, and derivative gains, respectively. During training, 
we back-propagate the gradient through the PID controller to the scanpath generator for parameter update.

\paragraph{\fontsize{12}{12}\selectfont{4. Implementation Details}}\mbox{}\\
\vspace{-4pt}

\noindent
The first stage of training is carried out by the Adam method~\cite{kingma2014adam} on  VRVQW  with an initial learning rate of $10^{-4}$ and a minibatch size of $48$. After the $50$-th epoch, the learning rate decays by a ratio of $0.1$, and we pre-train the scanpath generator for a total of $100$ epochs. The parameters for the PID controller are determined using the Ziegler–Nichols method~\cite{ziegler1942optimum}. 

For the second and third stages of training, the Adam method is also employed, and the detailed settings of different quality assessors are as follows.

\noindent\textbf{ScanpathVQA}. In the second training stage, the quality assessor is trained for $30$ epochs, with an initial learning rate of $5 \times 10^{-5}$, a decay ratio of $0.95$ per $2$ epochs, and a batch size of $8$. In the third stage of training on VRVQW, the entire method is trained for $5$ epochs, with an initial learning rate of $10^{-6}$, a decay ratio of $0.1$ after the $2$-nd epoch, and a batch size of $4$.  In the third stage of training on CVIQD and OIQA, the method is trained for $5$ epochs, with an initial learning rate of $10^{-5}$, a decay ratio of $0.9$ per epoch, and a batch size of $4$.

\noindent\textbf{GSR-S/GSR-X}.  In the second training stage, we follow the settings described in the original paper~\cite{sui2023perceptual}. In the third stage, the entire method is trained for $10$ epochs, with an initial learning rate of $10^{-6}$, a decay ratio of $0.9$ per $2$ epochs, and a batch size of $8$.  The input configuration also follows the original paper~\cite{sui2023perceptual}.

\noindent\textbf{Assessor360}. The settings in the second training stage are identical to those of ScanpathVQA. In the third stage, the entire method is trained for $10$ epochs, with an initial learning rate of $10^{-5}$, a decay ratio of $0.9$ per $2$ epochs, and a batch size of $4$. Owing to limitations in computer memory, the number of viewport sequences $N$ is reduced from $20$ to $15$.

\end{document}